# Stability and superconductivity of freestanding two-dimensional transition metal boridene: $M_{4/3}B_2$

Xiaoran Shi[1], Junfeng Gao[1*], Shi Qiu[1], Yuan Chang[1], Luneng Zhao[1], Zhen-Guo Fu[2*] Jijun Zhao[1] and Ping Zhang[2].

[1] Key Laboratory of Materials Modification by Laser, Ion and Electron Beams, Dalian University of Technology, Ministry of Education, Dalian 116024, China

[2] Institute of Applied Physics and Computational Mathematics, Beijing 100088, China

* Author to whom any correspondence should be addressed.

**E-mail:** gaojf@dlut.edu.cn, fu_zhenguo@iapcm.ac.cn

## Abstract

The small atomic mass of boron indicates strong electron-phonon coupling, so it may have a brilliant performance in superconductivity. Recently, a new 2D boride sheet with ordered metal vacancies and surface terminals ($Mo_{4/3}B_{2-x}$) was realized in experiments (*Science 2021, 373, 801*). Here, the 2D monolayer freestanding $Mo_{4/3}B_2$ is evidenced to be thermodynamically stable. Through electronic structure, phonon spectrum and electron-phonon coupling, monolayer $Mo_{4/3}B_2$ is found to be an intrinsic phonon-mediated superconductor. The superconducting transition temperature ($T_c$) is determined to be 4.06 K by the McMillian-Allen-Dynes formula. Remarkably, the $T_c$ of monolayer $Mo_{4/3}B_2$ can be increased to 6.78 K with an appropriate biaxial tensile strain (+5%). Moreover, we predict that other transition metal replacing Mo atoms is also stable and retaining the superconductivity. Such as monolayer $W_{4/3}B_2$ is also a superconductor with the $T_c$ of 2.37 K. Our research results enrich the database of 2D monolayer superconductors and boron-related formed materials science.

structure

# 1. Introduction

According to conventional Bardeen-Cooper-Schrieffer (BCS) theory[1], high Debye temperature and small atomic mass of light elements indicate strong electron-phonon coupling, which in turn can realize superconductivity. Therefore, hydrogen with the smallest atomic mass may theoretically have a higher superconducting transition temperature. Later, 'metal hydrogen', which can be converted from solid hydrogen under extremely high pressure, is indeed proved to be high temperature superconductor[2-7]. Due to the limitation of experimental conditions[8-9], researchers have turned their attention to other light element compounds, such as metallized hydrogen rich materials[10-15], which can achieve high temperature superconductivity under relatively low pressure. As the fifth element in periodic table, boron is the most promising element to realize superconductivity except hydrogen.

In fact, as early as 2001, the $T_c$ of bulk $MgB_2$ was predicted to reach ~40 K by BCS theory[16-17]. Such a high $T_c$ is mainly controlled by the B atom in the material through the electron in the B plane, which leads to the appearance of $\sigma$- and $\pi$– gap, is coupled to the lattice vibration of the B atom[18-22]. As one hotspot in condensed matter physics, introducing superconductivity in two-dimensional (2D) materials becomes an interesting issue[23-25]. As an illustrative example, borophene[26-27], a monolayer material of boron, have been successfully synthesized in experiments. Due to the electron deficiency, boron sheets show hexagonal holes or bulges in the hollow triangular sheets[28]. However, the boron based graphene-like honeycomb materials cannot exist stably. In particular, the variable structure causes them to exhibit metallic or semi metallic properties[29-30]. The study reveals that borophene may be the material with the highest $T_c$ among the current pure elemental 2D materials, whose $T_c$ is expected to be between 10-20 K[28]. The $T_c$ of 8-C2/m-II boron structure, a double-layer borophene sheets, can even reach 27.6 K[31]. The success of $MgB_2$ and borophene has confirm that boron can induce superconductivity.

Up to now, the borides has gradually developed into the largest family of light element superconductors[32]. Significantly, superconductivity as low as single layer thickness is also realized in $MgB_2$[33-34], which can be further boosted to above 50 K by merely ~4% of tensile biaxial strain. In addition, Song et al.[35] predicted a 2D aluminum boride ($AlB_6$–ptAl–array) nanosheet with Tc calculated as 4.7K. Bo et al.[36] discussed the possibility of superconducting in $MB_6$ (M=Mg, Ca, Sc, Ti, Y). In addition, Modak et al.[37] reported that LiBC exhibited a $T_c$ of 70 K. This further inspired us to study the possible superconductivity in new 2D boride materials.

Recently, by selective etching of metal atoms from $(Mo_{2/3}Y_{1/3})_2AlB_2$ and $(Mo_{2/3}Sc_{1/3})_2AlB_2$ in aqueous hydrofluoric acid, Zhou and collaborators have experimentally achieved the 2D boride, a real single layer $Mo_{4/3}B_{2-x}T_z$[38]. The pristine 2D boride without surface termination is supposed to be $Mo_{4/3}B_2$. Considering the importance in both the perspectives of basis physics and potential applications of 2D borides in superconductivity electronics fields, it is interesting and timely necessary to carry out a theoretical analysis of the physical properties.

In this work, we use many-body theory combined with first-principles calculations to study electronic properties and electron-phonon coupling (EPC) of the bare 2D monolayer $Mo_{4/3}B_2$ is not only a metallic monolayer but also an inherent superconductor with the $T_c$ of 4.06 K. Inspiringly, the value of $T_c$ can be effectively increased to 6.78K by applying biaxial strain, Extensively, we indicate rich 2D monolayer boride superconductor with $M_{4/3}B_2$ (M is transition metal) formation also can be achieved by element replacing, such as monolayer $W_{4/3}B_2$.

## 2. Results and discussions

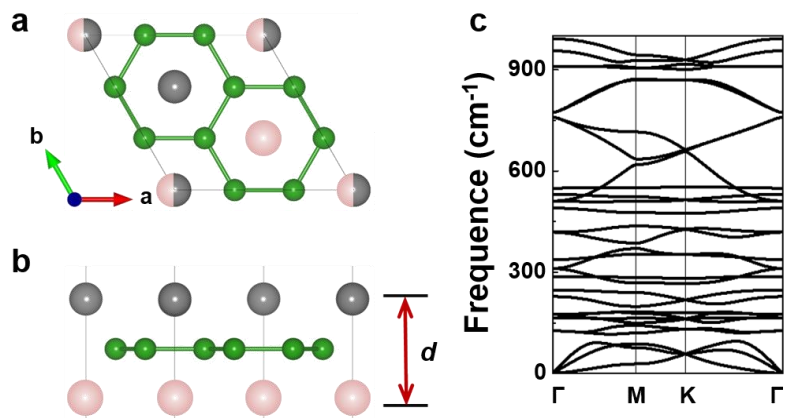

**Figure 1.** (Color online) (a) Top and (b) side views of Monolayer $Mo_{4/3}B_2$. The green balls indicate B atoms while the gray and pink balls indicate the atoms of top and bottom Mo-layers, respectively. The atoms in the overlapping parts of the two layers are colored half gray and half pink. The unit cell is indicated by the solid black line. (c) Phonon spectrum of the pristine $Mo_{4/3}B_2$ monolayer.

As shown in Figure 1, $Mo_{4/3}B_2$ is a layered compound with a hexagonal *P-3m1* (no.164) space group. The three sublayers are stacked in the sequence of Mo-B-Mo. The thickness of $Mo_{4/3}B_2$ monolayer as indicated by *d* in Figure 1b is 2.89 Å. After full relaxation, the lattice constant is 5.14 Å. Figure 1c depicts the phonon dispersion along the Γ-M-K-Γ path, clearly showing that the 2D monolayer $Mo_{4/3}B_2$ is dynamically stable due to the absence of imaginary phonon mode. A series of biaxial strains are applied to $Mo_{4/3}B_2$ monolayer to check its stability under strain. According to the phonon dispersion shown in Figure S1, we find that the monolayer $Mo_{4/3}B_2$ keeps dynamically stable under the biaxial strain of $-1\% < \varepsilon < 7\%$. By comparing phonon band structures under tensile strain, it can be seen that the interatomic charge density between atoms is depleted as the interatomic distance increases, resulting in softer mode of phonons.

To analyze the bonding characteristics of the structure, the charge density difference is calculated and shown in Figure S2. There are considerable charge redistributions between the B monlayer and metal atoms, indicating relatively strong interactions. According to Barder analysis, 0.4 *e* transferred from Mo atoms to the B atoms, that is why the honeycomb monolayer of boron can be stabilized.

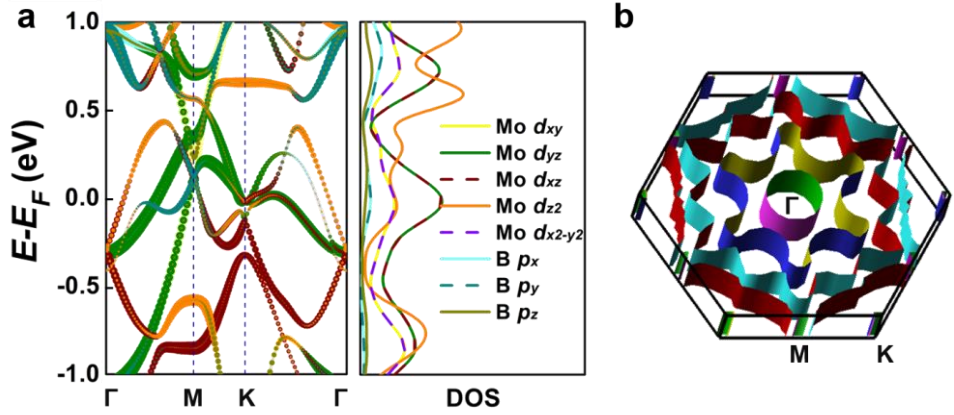

**Figure 2.** (Color online) (a) Orbital-projected band structures and densities of states of $Mo_{4/3}B_2$. The Fermi level is set to zero. (b) Fermi surface in the first Brillouin zone.

The orbital-resolved band structures and electronic density of states (DOS) of $Mo_{4/3}B_2$ are presented in Figure 2a. The Fermi surface is shown in Figure 2b. It can be seen that $Mo_{4/3}B_2$ exhibits inherent metallic properties with several partially occupied bands crossing over the Fermi level. There are four, three, and five bands crossing the Fermi level along Γ-M, M-K and K-Γ high-symmetry paths, respectively, which are consistent with the Fermi surface, which indicating that it is a good electronic conductor. As shown in Figure 2b, these bands form one circular and two polygonal star-shaped electron pockets centered at Γ and a triangular-like electron pocket centered at K. Moreover, the bands around the Fermi level are mainly dominated by Mo-$d$ orbitals, especially $d_{xz}$ and $d_{yz}$ orbitals, and the contribution from the B-$p$ orbital is minor.

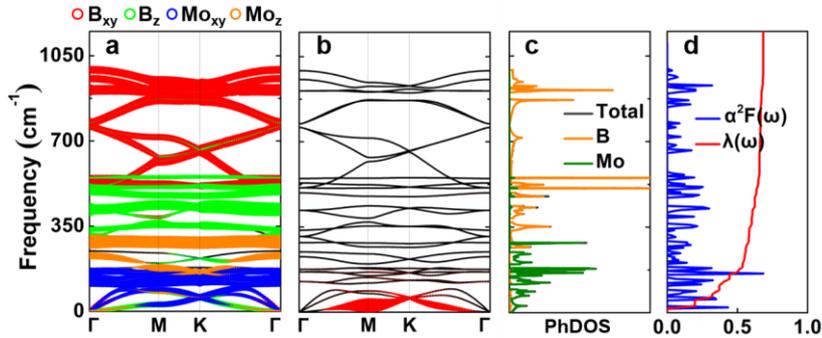

**Figure 3.** (Color online) Phonon dispersions, PhDOS, Eliashberg function $α^2F(ω)$ and integrated electron–phonon coupling strength $λ(ω)$ of $Mo_{4/3}B_2$. The red, green, blue, and orange hollow circles in (a) represent B horizontal, B vertical, Mo horizontal, and Mo vertical modes,

respectively. The magnitude of $\lambda_{qv}$ is displayed with an identical scale in all figures for comparison. The size of red dots in (b) drawn proportional to the magnitude of $\lambda_{qv}$.

Next, we turn our attention to the EPC and superconducting transition temperature ($T_c$) of Mo$_{4/3}$B$_2$. The phonon dispersion, magnitude of EPC $\lambda_{qv}$, phonon density of states (PhDOS), Eliashberg electron–phonon spectral function $\alpha^2F(\omega)$, and cumulative frequency dependent EPC $\lambda(\omega)$ are presented in Figure 3. The magnitude of EPC $\lambda_{qv}$ can be calculated from the Migdal-Eliashberg equation[39-40] according to the BCS theory,

$$\lambda_{qv} = \frac{\gamma_{qv}}{\pi h N(E_F) \omega_{qv}^2}, \tag{1}$$

where $\gamma_{qv}$ is the phonon linewidth, $\omega_{qv}$ is the phonon frequency, and $N(E_F)$ is the electronic DOS at the Fermi level. We can estimate $\gamma_{qv}$ by

$$\gamma_{qv} = \frac{2\pi\omega_{qv}}{\Omega_{Bz}} \sum_{k,n,m} \left| g^v_{kn,k+qm} \right|^2 \delta(\varepsilon_{kn} - \varepsilon_F)\delta(\varepsilon_{k+qm} - \varepsilon_F), \tag{2}$$

where $\Omega_{BZ}$ is the volume of the Brillouin zone (BZ), $\varepsilon_{kn}$ and $\varepsilon_{k+qm}$ represent the Kohn–Sham energy, $\varepsilon_F$ is the Fermi energy, and $g^v_{kn,k+qm}$ denotes the EPC matrix element that can be determined self-consistently by the linear response theory[41]. The Eliashberg electron–phonon spectral function $\alpha^2F(\omega)$ and the total EPC constant $\lambda$ are estimated by

$$\alpha^2F(\omega) = \frac{1}{2\pi N(E_F)} \sum_{qv} \frac{\gamma_{qv}}{\omega_{qv}} \delta(\omega - \omega_{qv}), \tag{3}$$

and

$$\lambda(\omega) = 2 \int_0^\omega \frac{\alpha^2F(\omega)}{\omega} d\omega, \tag{4}$$

respectively. The logarithmic average frequency $\omega_{\log}$ is expessed as

$$\omega_{log} = \exp\left[\frac{2}{\lambda} \int_0^\infty \frac{d\omega}{\omega} \alpha^2F(\omega) \log \omega \right]. \tag{5}$$

Then, the superconducting transition temperature $T_c$ can be evaluated using the Allen–Dynes[41-43] modified McMillan formula[44]

$$T_c = \frac{\omega_{\log}}{1.2} \exp\left[-\frac{1.04(1+\lambda)}{\lambda - \mu^*(1+0.62\lambda)}\right], \tag{6}$$

where an effective screened Coulomb repulsion constant $\mu^* = 0.1$ is used[44-46].

From the decomposition of the phonon spectrum relative to the vibration of Mo and B atoms and the PhDOS, one can clearly find that the vibration of Mo atoms mainly contributes to the low-frequency spectrum due to its heavy atomic mass. The in-plane vibration (Mo-xy) of the atoms dominates the region below 180 cm$^{-1}$, while the region from 180 to 310 cm$^{-1}$ is related entirely to the out-of-plane vibration (Mo-z). Besides, the vibration of the B atom appears almost in the entire frequency range. Among them, in the mid-frequency region from 310 to 550 cm$^{-1}$, the vibration mainly arises from the out-of-plane mode (B-z) of B atoms, and the modes above 550 cm$^{-1}$ are contributed from the in-plane vibration (B-xy) of the B atom. The highest phonon frequency is about 1000 cm$^{-1}$, which is much larger than that of $MoS_2$ (473 cm$^{-1}$)[47] and slightly smaller than that of borophene (1274 cm$^{-1}$)[48], indicating the strong bonding interaction between the B atoms.

As shown in Figure 3b, the low-frequency modes of phonon play key role to the realization of EPC in $Mo_{4/3}B_2$ monolayer, accounting for 86.03% of the total EPC. The Eliashberg function $\alpha^2 F(\omega)$ shows that the corresponding region has five main peaks located at 22.13, 55.31, 132.76, 160.41 and 182.54 cm$^{-1}$, respectively (see Figure 3d). In the frequency range below 34 cm$^{-1}$, $\lambda_{qv}$ along the Γ-M-K-Γ direction is most pronounced and leads to the first peak of $\alpha^2 F(\omega)$, which accounts for about 54.34% of the total EPC. As a consequence, $Mo_{4/3}B_2$ monolayer is a medium-coupling superconductor with $\lambda$ of 0.68 according to the role proposed by Allen *et al.*[32] and possesses a $T_c$ of 4.06 K.

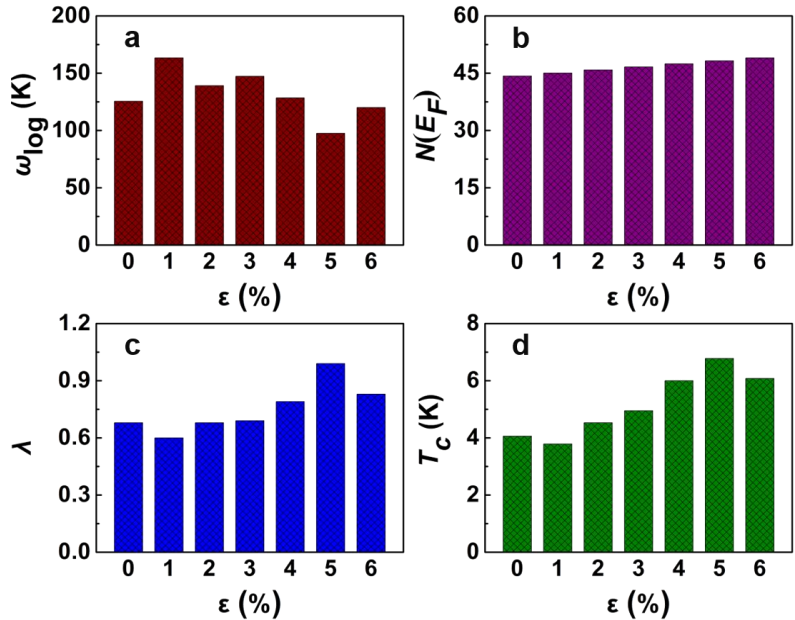

**Figure 4.** (Color online) The changes of the superconductive parameters of (a) $\omega_{\log}$, (b) $N(E_F)$, (c) $\lambda$, and (d) $T_c$ as a function of the strength of tensile strain ε.

Furthermore, we investigate the potential effect of strain on the superconductivity of monolayer $Mo_{4/3}B_2$. Figure 4 shows the variations of $\omega_{log}$, $N(E_F)$, $\lambda$, and $T_c$ under a series of tensile strains. As the tensile strain increases, $\omega_{log}$ first increases until the strain is greater than 1%, and then gradually decrease. $N(E_F)$ monotonically increases with the strain increasing. $\lambda$ and $T_c$ emerge the same trend of change, which roughly negatively correlated with the change of $\omega_{log}$. Remarkably, the $T_c$ of $Mo_{4/3}B_2$ with 5% tensile strain can reach the maximum value of 6.78 K with an EPC constant $\lambda = 0.99$, which is significantly higher than the temperature of liquid helium (4.2 K).

In addition, with the further increase of the tensile strain, the $T_c$ decreases slowly. However, when the strain exceeds 7%, imaginary frequencies will occur along the M-K direction in the phonon spectrum (see Figure S1), which suggest that the $Mo_{4/3}B_2$ monolayer may be tuned into a new charge density wave phase. In this case, the charge density wave should compete or coexist with the superconductivity. Namely, the superconductivity may disappear or merge into a possible charge density wave state. In order to illustrate this issue, the corresponding calculations need to be carried out under a reasonable charge density wave structure, which is beyond the scope of this paper. The results of electronic structures and EPC make clear that the superconductivity in monolayer $Mo_{4/3}B_2$ is dominated by the $d$-orbital of Mo, and tensile strain induced soft-mode phonons along M-K direction (especially at K point) is crucial to tunable of the superconductivity. Consequently, the results shown here indicate that tensile strain engineering can be used as an effective method to tune the superconductivity in monolayer $Mo_{4/3}B_2$.

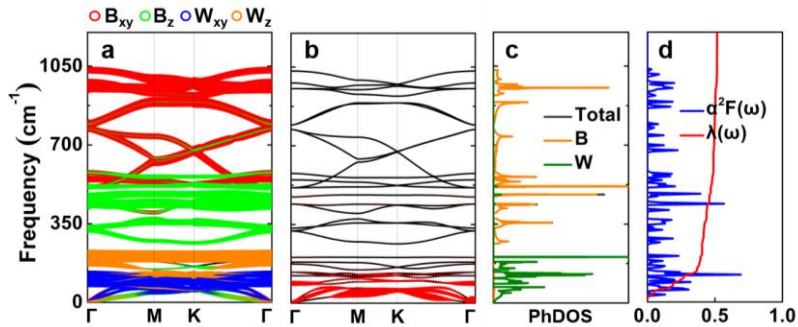

**Figure 5.** (Color online) Phonon dispersions, PhDOS, Eliashberg function $\alpha^2F(\omega)$ and integrated electron–phonon coupling strength $\lambda(\omega)$ of monolayer $W_{4/3}B_2$. The red, green, blue and orange

hollow circles in (a) represent B horizontal, B vertical, W horizontal and W vertical modes, respectively. The magnitude of $\lambda_{qv}$ is displayed with an identical scale in all figures for comparison. The size of red dots in (b) drawn proportional to the magnitude of $\lambda_{qv}$.

Inspired by the success of replacing metal atoms in transition metal chalcogenides[49], we construct monolayer $W_{4/3}B_2$ by replacing Mo atoms in $Mo_{4/3}B_2$ by W atoms. As shown in Figure 5, $W_{4/3}B_2$ is also dynamically stable. As seen from the decomposed phonon dispersion and PhDOS of Figure 5, the entire range of phonon spectrum is divided into low-frequency and high-frequency regions by an obvious phonon gap. Compared with $Mo_{4/3}B_2$, the coupling between B-B in $W_{4/3}B_2$ is stronger, and the highest phonon frequency can reach 1050 cm$^{-1}$. The in-plane B-xy vibrations dominate high frequencies above 513 cm$^{-1}$, while the out-of-plane Bz vibrations dominate the intermediate frequencies from 265 cm$^{-1}$ to 513 cm$^{-1}$. The W atoms dominate the low frequencies below 205 cm$^{-1}$, and the in-plane W-xy vibration, which dominates the region below 140cm$^{-1}$, contributes the most to the EPC, accounting for 78.57 % of the total EPC. Similarly, relatively large values of λqv along the Γ–M–K–Γ directions are found, but they are obviously smaller than that of $Mo_{4/3}B_2$. Therefore, the value of $T_c$ for $W_{4/3}B_2$ is lower than that for $Mo_{4/3}B_2$, which is 2.37 K with λ = 0.51.

## 3. Conclusions

In summary, starting from the recently discovered 2D boridene $Mo_{4/3}B_{2-x}T_z$, we have studied the electronic structure and electron-phonon coupling of the perfect monolayer through first-principles calculations. The results indicate that $Mo_{4/3}B_2$ with high stability is an intrinsic conventional superconducting material with a critical temperature $T_c$ =4.06 K, where the superconductivity can be significantly enhanced by tensile strain. When 5% tensile strain is applied, $T_c$ can reach a maximum value of 6.78 K. Furthermore, it may be a potential phonon topological material. Finally, we predict that the $W_{4/3}B_2$ obtained by replacing the metal atoms of the same group is also a superconductor with $T_c$ =2.37 K. The above results will further trigger the efforts of 2D superconducting materials.

## Acknowledgements

This work is supported by the National Natural Science Foundation of China (Grant No. 12074053, 91961204, 12004064), by the Fundamental Research Funds for the Central Universities (DUT21LAB112, DUT22ZD103, DUT22LK11). We also acknowledge Computers supporting from Shanghai Supercomputer Center, DUT supercomputing center, and Tianhe supercomputer of Tianjin center.